\documentclass{aastex6}
\usepackage[fleqn]{amsmath}
\usepackage{bm}

\fullcollaborationName{The Friends of AASTeX Collaboration}

\begin{document}

\title{Formation of Galactic Prominence in Galactic Central Region}


\author{Chih-Han Peng\altaffilmark{1} and Ryoji Matsumoto\altaffilmark{1}}
\affil{\altaffilmark{1}Department of Physics, Graduate School of Science, Chiba University,  \\
1-33 Yayoi-cho, Inage-ku, Chiba 263-8522, Japan}






\altaffiltext{1}{Department of Physics, Graduate School of Science, Chiba University; julian8128@chiba-u.jp}

\begin{abstract}

 We carried out 2.5-dimensional resistive MHD simulations to study the formation mechanism of molecular loops observed by \citet{2006Sci...314..106F} at Galactic central region. Since it is hard to form molecular loops by uplifting dense molecular gas, we study the formation mechanism of molecular gas in rising magnetic arcades. This model is based on the in-situ formation model of solar prominences, in which prominences are formed by cooling instability in helical magnetic flux ropes formed by imposing converging and shearing motion at footpoints of the magnetic arch anchored to the solar surface. We extended this model to Galactic center scale (a few hundreds pc). Numerical results indicate that magnetic reconnection taking place in the current sheet formed inside the rising magnetic arcade creates dense blobs confined by the rising helical magnetic flux ropes. Thermal instability taking place in the flux ropes forms dense molecular filaments floating at high Galactic latitude. The mass of the filament increases with time, and can exceed 10${^5}$ M$_{\odot}$.

\end{abstract}




\section{Introduction} \label{sec:intro}

Molecular loops are found in the Galactic central region by NANTEN CO observations (\citealt{2006Sci...314..106F,2009PASJ...61.1039F}, see \citealt{2010PASJ...62..675T, 2010PASJ...62.1307T} and \citealt{2011PASJ...63..171K} for detailed observations of these molecular loops). \citet{2006Sci...314..106F} proposed that Parker instability \citep{1966ApJ...145..811P} is a possible mechanism to create these molecular loops. This model is motivated by the systematic variation of the line of sight velocity along the loop; the velocity distribution can be explained by the motion of the dense gas sliding down along the rising magnetic loops formed by the Parker instability. 
Furthermore, large velocity dispersion observed around the footpoints of the loop can be explained by shock waves formed in the region where the supersonic downflow hits the galactic gas disk \citep{1988PASJ...40..171M,1990ApJ...356..259M,1991Natur.353..633S,2009PASJ...61..411M,2009PASJ...61..957T}. Nevertheless, it is a puzzle how dense molecular gas (number density $>$ 100 cm$^{-3}$) can be levitated in warm interstellar medium (number density = a few cm$^{-3}$). The levitated dense molecular filaments are similar to the solar prominences whose density is 10-100 times larger than the surrounding solar corona. Note that solar prominences are warm gas with temperature 10$^4$ K, meanwhile molecular loops are cold molecular gas with temperature less than 100K. Although the temperature and cooling mechanisms of solar prominences differ from those in molecular loops, the density ratio between the dense filament and ambient gas is similar. \citet{2010PASJ...62.1307T} discussed the possibility that the formation mechanism of molecular loops is similar to solar prominences. Understanding the physical process of the formation of solar prominences might be a key to reveal the formation mechanism of molecular loops.

  Recently, \citet{2015ApJ...806..115K} proposed a model of in-situ formation of solar prominence by condensation of hot coronal gas. They showed that magnetic flux ropes can be formed by converge/shear motion for magnetic arches anchored to the solar photosphere. Dense plasma trapped in the flux rope are condensed by radiative cooling and can form cool, dense filaments observed as prominences. 
  
In this paper, we present the results of 2.5-dimensional MHD simulations based on the model of solar prominences by \citet{2015ApJ...806..115K}. Cooling/heating function is adopted from \citet{2006ApJ...652.1331I}, in which radiative cooling by emission lines, photoelectric effect on dust, cosmic rays, etc. are included. We demonstrate that dense loop-like structures near the Galactic center can be formed by condensation of the warm gas. 

  Our model and numerical settings are introduced in Section 2. Section 3 shows our numerical results. We compare the numerical results with observations. Section 4 is for summary and discussion.
\section{Numerical Model} \label{sec:Numerical Method}

\subsection{In-situ Formation Model of Galactic Prominence} \label{subsec:tables}
We adopted the in-situ formation model of prominences reported by \citet{2015ApJ...806..115K}. Figure 1 schematically shows the formation mechanism of dense, cold filaments. We consider static magnetic arches at the initial state Figure 1(a)). Subsequently, we impose converging and shearing motion at the footpoints of the arch to induce the rising motion of the arcade. Note that both the magnetic buoyancy around the loop top of the arch and the magnetic pressure gradient enhanced by the footpoint motion contribute to the rising motion of the arch. Current sheets can be formed inside the expanding magnetic  arcade. Magnetic reconnection taking place in the current sheet forms rising dense plasmoids (Figure 1(b)). Since the plasmoid is confined by the helical magnetic fields, the warm gas accumulates around the bottom of the closed poloidal magnetic field lines (red arrows), where the cooling instability forms dense, cold filaments supported by magnetic tension force (Figure 1(c)). In this paper, since we neglect the variation of physical quantities perpendicular to the poloidal plane, the dense filaments are not arches but straight filaments. In three-dimension, the filaments can bend and form an arch, along which the molecular gas slides down.

\begin{figure}
\figurenum{1}
\plotone{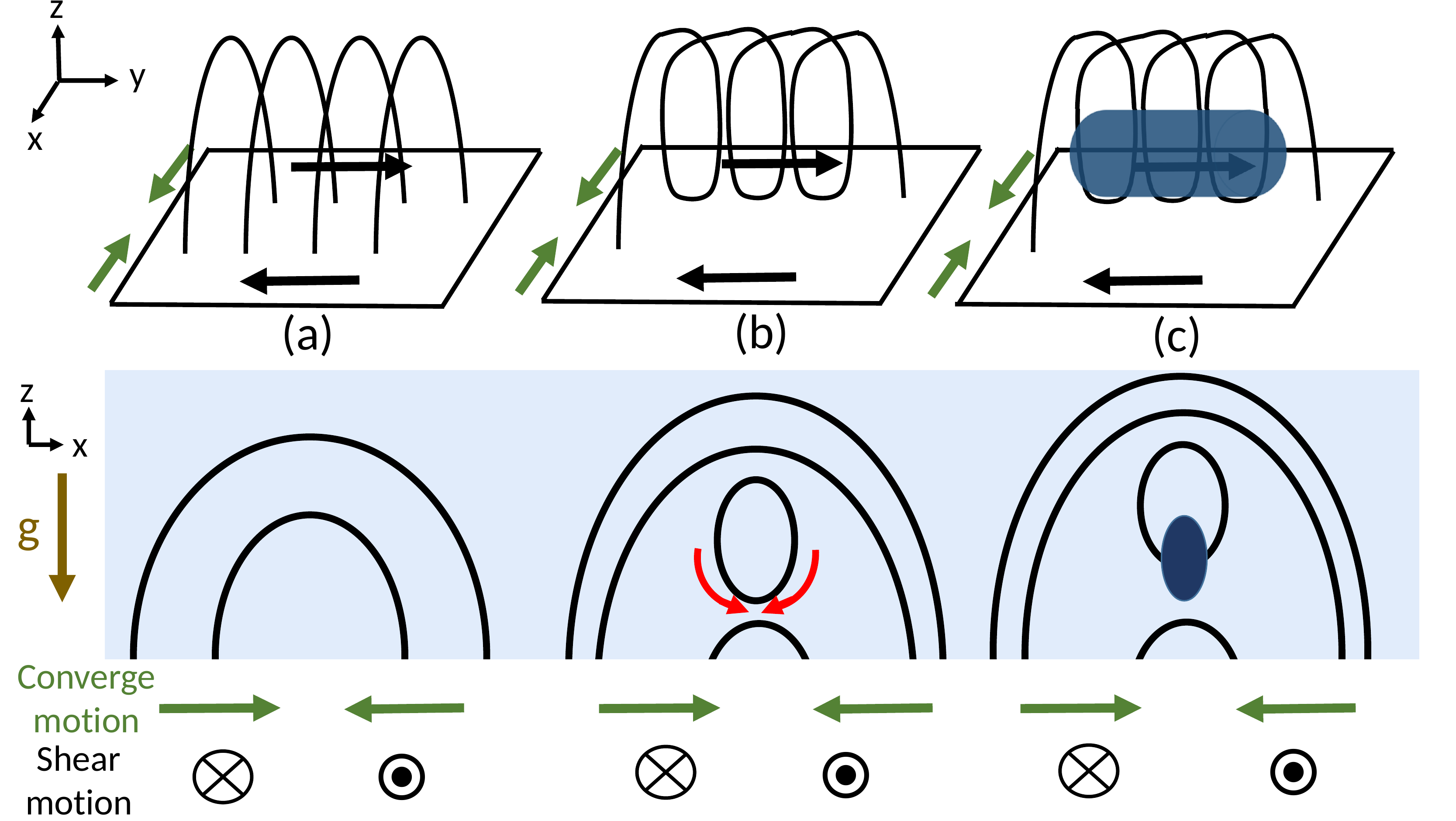}
\caption{Schematic picture of molecular loop formation. (a) Magnetic arches and footpoint motions (b) Closed magnetic loops (helical magnetic fields in three-dimension) are formed by magnetic reconnection taking place in the arch expanding by the imposed converging/shear motion. (c) Cooling instability taking place in the flux rope forms dense filaments (dark region).\label{fig:model}}
\end{figure}
\subsection{Basic Equations and Numerical Scheme} \label{subsec:tables}
The basic equations of resistive magnetohydrodynamics (MHD) are as follows:
\begin{flalign}
&\frac{\partial \rho}{\partial t} +\nabla\cdot(\rho \bm{v})=0,\\ 
&\frac{\partial (\rho\bm{v})}{\partial t} +\nabla\cdot(\rho \bm{v}\bm{v})=-\nabla P-\rho\bm{g}+\frac{(\nabla\times\bm{B})\times\bm{B}}{4\pi},\\
&\frac{\partial \bm{B}}{\partial t} =\nabla\times(\bm{v} \times\bm{B}-\eta \nabla \times \bm{B}),\\
&\frac{\partial E}{\partial t} +\nabla\cdot\bigg[\Big(E+P+\frac{B^{2}}{8\pi}\Big)\bm{v}-\frac{\bm{B}(\bm{B}\cdot\bm{v})-\eta(\nabla \times \bm{B})\times \bm{B}}{4\pi}\bigg]=\rho\bm{v}\cdot\bm{g}-\rho\mathcal{L},\\
&\qquad E=\frac{P}{\gamma-1}+\frac{\rho v^{2}}{2}+\frac{B^{2}}{8\pi}.
\end{flalign}
where the anomalous resistivity $\eta$ is assumed to depend on the current density $J$ (see \citealt{2015ApJ...806..115K}) as
\begin{flalign}
&\qquad \eta=\begin{cases}
          0 & J < J_c,\\
          \eta_0(J/J_c -1)^2  & J \geq J_c.
     \end{cases}
\end{flalign}
Here $\eta_0$ = 3 $\times$ 10$^{23}$ cm$^{2}$ s$^{-1}$ and $J_c$ = 4.0 $\times$ 10$^{-17}$ dyn$^{1/2}$ cm s$^{-1}$. We adopt Cartesian coordinates $(x,y,z)$ and assume translational symmetry with respect to $y$. The gravitational acceleration is assumed to be $\bm{g} = (0,0,g_z)$ and $g_z$ = 3 $\times$ 10$^{-9}$ cm s$^{-2}$. Cooling function $\rho\mathcal{L}$ includes cooling and heating term and other symbols have their normal meaning. In the interstellar medium, cold neutral medium with temperature 10-100 K and warm medium with temperature 10$^4$ K coexists by the balance of cooling and heating \citep{1969ApJ...155L.149F,1995ApJ...443..152W,2003ApJ...587..278W}. We adopted cooling/heating function summarized by \citet{2006ApJ...652.1331I}.    
\begin{flalign}
&\rho\mathcal{L}=(-\Gamma + n\Lambda)n,\\
&\Gamma = 2\times10^{-26} \ \text{erg} \ \text{s}^{-1},\\
&\Lambda = 7.3\times10^{-21}\ \text{exp}\Big(\frac{-118400}{T+1500}\Big)\nonumber \\
&+7.9\times10^{-27}\ \text{exp}(-92/T)\ \text{erg} \ \text{cm}^{3} \ \text{s}^{-1}.
\end{flalign}
Here $n$ is number density. Figure 2 shows a thermal equilibrium curve for the interstellar medium determined by the cooling/heating function (7)-(9). Two stable branches, warm medium (WM, red branch) and cold neutral medium (CNM, blue branch) exist. In this paper we assume that cooling/heating takes place at temperature range 400 K $< T <$ 20000 K, otherwise we set $\rho\mathcal{L} = 0$ in order for the dense, cold region be resolved numerically with moderate number of grid points. For simplicity, we neglect thermal conduction.

The basic equations are solved by applying the HLLD scheme \citep{2005JCoPh.208..315M}. Second order accuracy in space is preserved by linearly interpolating the values at the cell interface, and restricting them using the minmod limiter. The solenoidal condition $\nabla \cdot \bm{B}$ = 0 is satisfied by applying the generalized Lagrange multiplier (GLM) scheme proposed by \citet{2002JCoPh.175..645D}. The cooling/heating term is included by time-implicit method. This simulation code has been applied to MHD simulations of the interaction of jets with interstellar gas \citep{2014ApJ...789...79A}. 

The size of the simulation box is -200 pc $< x <$ 200 pc, and 5 pc $< z <$ 2400 pc. The grid size is uniform with $\Delta$ $x$ = 1 pc and $\Delta$ $z$ = 0.5 pc.

\begin{figure}
\figurenum{2}
\plotone{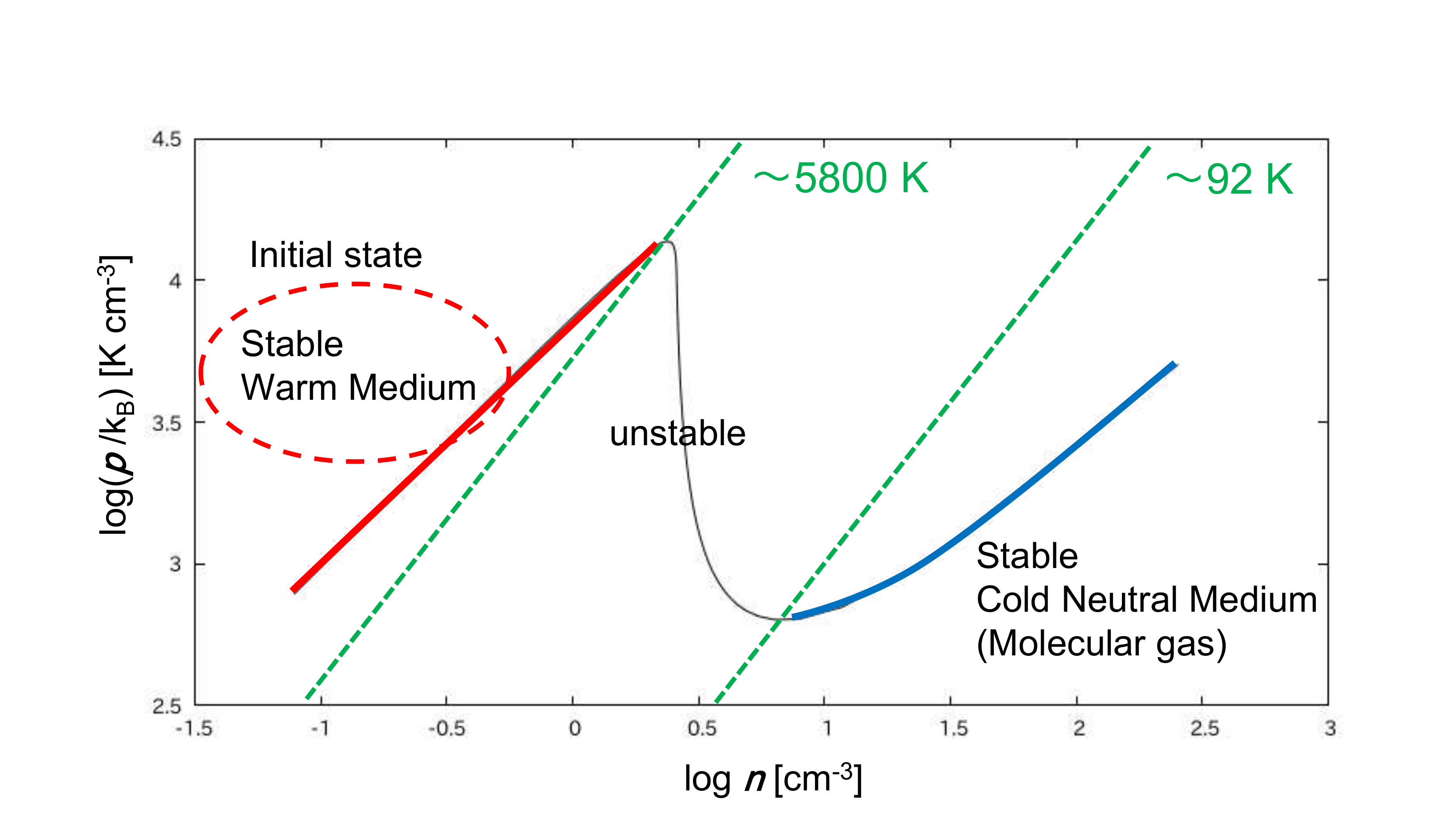}
\caption{Thermal equilibrium curve computed from the cooling/heating function adopted in this paper.\label{fig:thermal}}
\end{figure}
\subsection{Initial and Boundary Conditions} \label{subsec:tables}
We assume gravitationally stratified layer in hydrostatic equilibrium.  Initial magnetic field is assumed to be force-free and given by
\begin{flalign}
&B_x=-\Big(\frac{2L_a}{\pi H_m}\Big)B_a\cos  \Big(\frac{\pi}{2L_a}x\Big)\exp\Big(-\frac{z}{H_m}\Big),\\
&B_y= \sqrt{1-\Big(\frac{2L_a}{\pi H_m}\Big)^2}B_a\cos\Big(\frac{\pi}{2L_a}x\Big)\exp\Big(-\frac{z}{H_m}\Big),\\
&B_z= B_a\sin\Big(\frac{\pi}{2L_a}x\Big)\exp\Big(-\frac{z}{H_m}\Big).
\end{flalign}

Initial distribution of density, temperature and magnetic field are shown in Figure 3. We set $\rho$($z$ = 5 pc) = 2.8 $\times$ 10$^{-24}$  g cm$^{-3}$ and $P$($z$ = 5 pc) = 1.56 $\times$ 10$^{-12}$  dyn cm$^{-2}$ ($T$($z$ = 5 pc) = 6700 K). Here $H_m$ = 200 pc and $L_a$ = 200 pc denote magnetic scale height and half width of magnetic arch, respectively. We assume strong magnetic field to sustain the dense filaments. The plasma $\beta$ ( = $P_{gas}$/$P_{mag}$) is assumed to be $\beta$ = 0.2 at bottom of the simulation area ($B_a$ = 1.54 x 10$^{-5}$ G). The warm gas with temperature 6700-20000 K is assumed to satisfy mechanical equilibrium and thermal equilibrium before converge and shear motions are imposed. In region where $z >$ 900 pc, we assume the hot corona with $T_{corona}$ = 2 $\times$ 10$^6$ K.

Boundary conditions in x-direction are symmetric for $\rho$, $P$, $B_z$, $v_z$ and anti-symmetric for $B_x$, $B_y$, $v_x$, $v_y$. Absorption boundary conditions are applied at the upper boundary. At the lower boundary, density and pressure are fixed to the initial value. Converge ($v_x$) and shear ($v_y$) motions are imposed at foot points of magnetic arch as follows : 
\begin{flalign}
&v_x=v_y=\begin{cases}
          -v_0\frac{ \displaystyle{x}}{ \displaystyle{L_a/4}} & 0 \leq x < L_a/4,\\
          -v_0\frac{ \displaystyle{L_a-x}}{ \displaystyle{3L_a/4}} & L_a/4 \leq x \leq L_a,
     \end{cases}\\
&v_z=0.
\end{flalign}
Here $v_0$ is 4 km/s
\begin{figure}
\figurenum{3}
\gridline{\fig{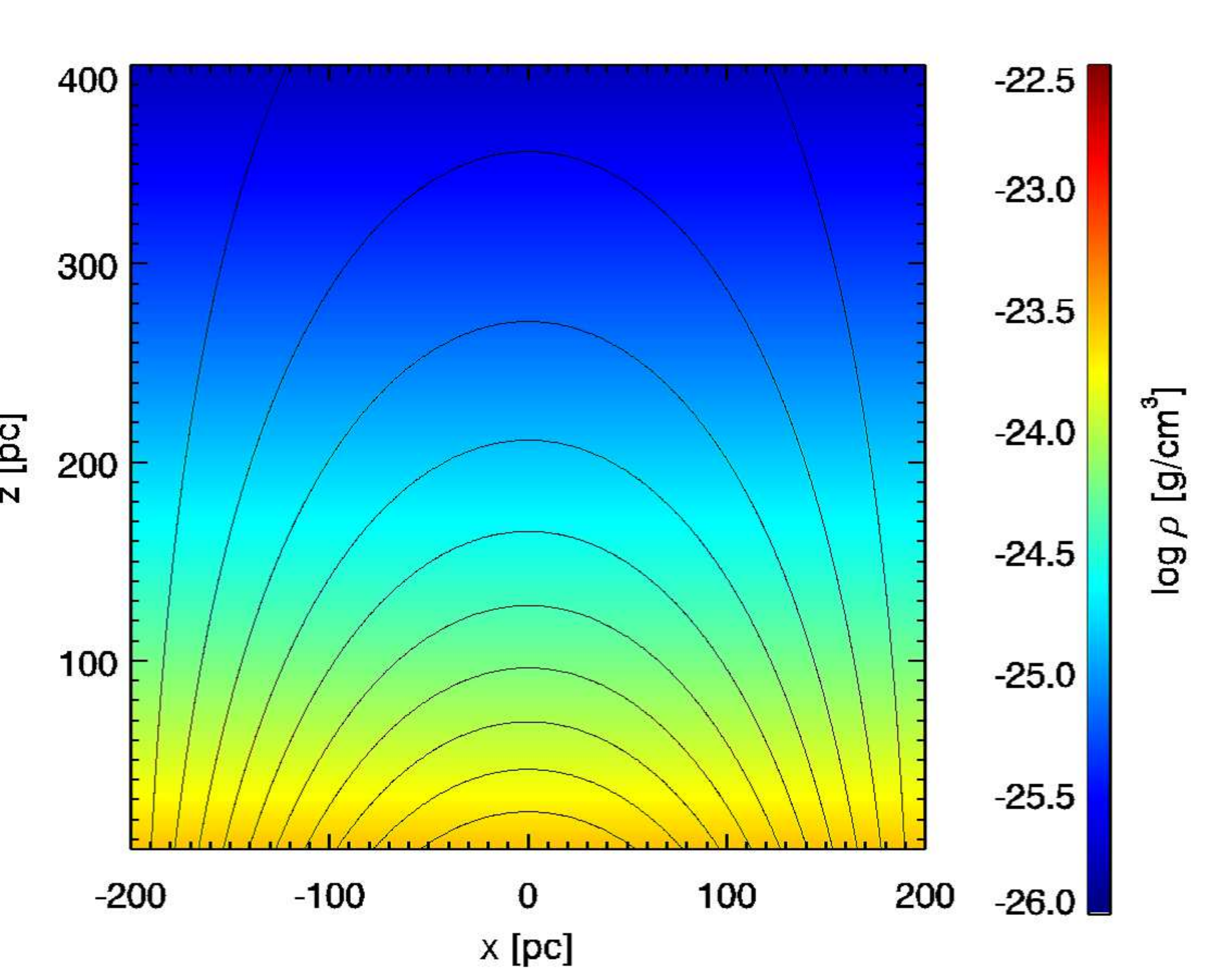}{0.45\textwidth}{}
          \fig{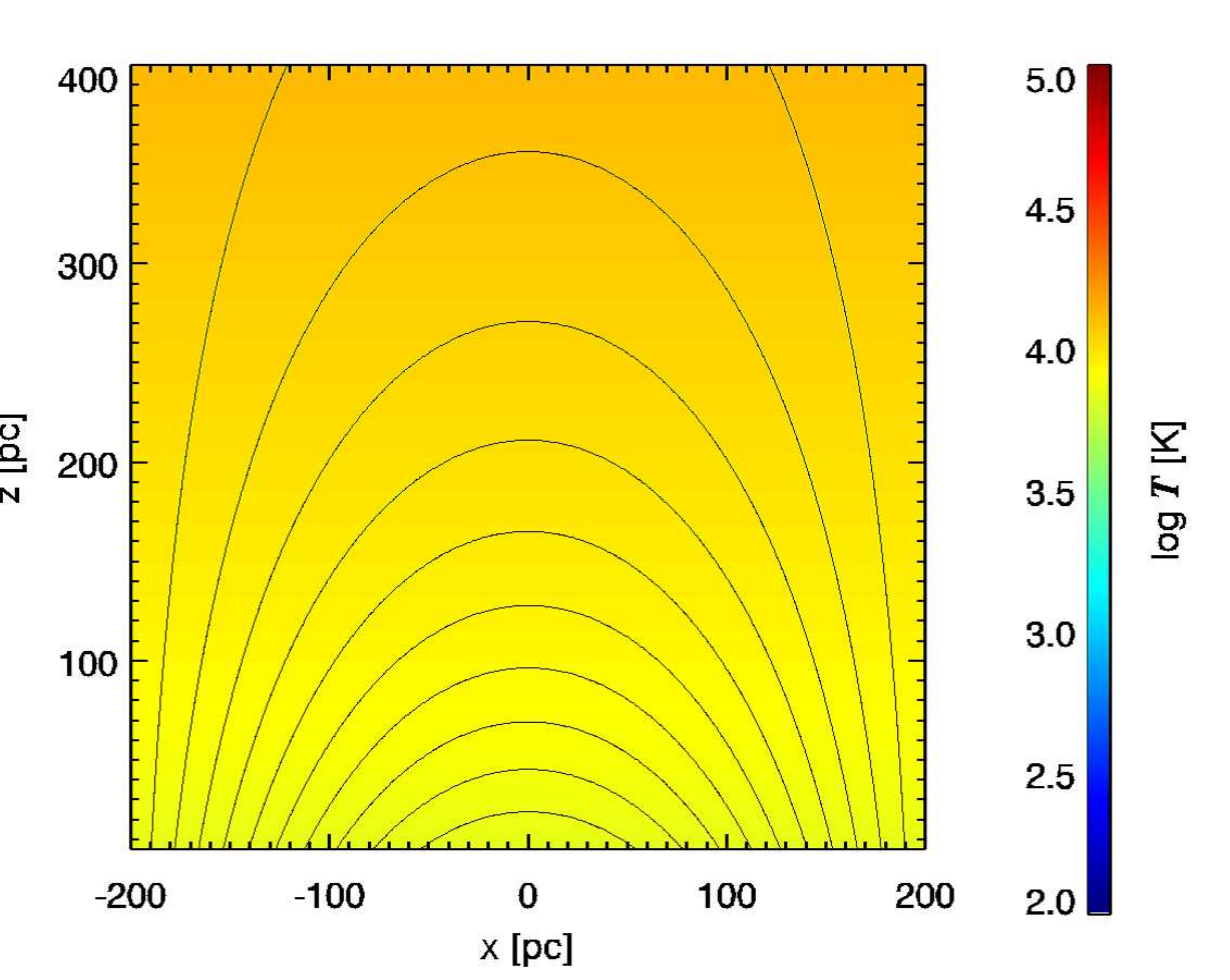}{0.45\textwidth}{}
          }
\caption{Inititial distribution of density (left), temperature (right) and magnetic field lines (solid curves).}
\end{figure}

\section{Numerical Results and Comparison with Observation} 
\subsection{Formation of A Dense Filament}
Figure 4 shows density and temperature distribution in $x$-$z$ plane at 30 Myrs, 40 Myrs and 120 Myrs. The arcade field squeezed by the converge motion and stretched by the shear motion form a current sheet around $x$ = 0 inside the arcade. Magnetic reconnection taking place in the current sheet creates rising flux ropes inside which warm gas is confined. In this simulation, the flux rope is lifted up to $\sim$250 pc. Since the warm interstellar gas slides down along the closed magnetic field lines, the density around the bottom of the flux rope exceeds the threshold for the onset of the cooling instability. Thermal instability taking place in the dense region forms cold, dense filaments. Density of the filament becomes 10-100 times that of the initial state. The dense filament is lifted up to 100-200 pc and sustained for 100 Myrs. The length and thickness of the filament at 120 Myrs are $\sim$60 pc in vertical direction and 6-8 pc in $x$-direction, respectively. 

Figure 5 enlarges the region where dense filaments are formed. Arrows show velocity vectors. The warm gas infalls along the magnetic field lines. The speed of the warm gas accumulating toward the filament is about 2 km/s. Figure 6 shows the distribution of $B_y$ at t = 120 Myrs. Since $B_y$ is amplified by shear motion, $B_y$ becomes strong inside the magnetic flux rope. The flux rope corresponds to the blue-purple region in Figure 6 where $B_y >$ 5.5 $\mu$G. Magnetic field lines inside the flux rope have helical shape as schematically shown in Figure 1(b). 

Red dots in Figure 7 show the pressure and number density at each grid point. At the initial state ( Figure 7(a)), warm gas locates on the thermal equilibrium curve (solid line). Dashed line shows the isothermal line at $T$ = 400 K and the dash-dotted line shows that at $T$ = 20000 K. Figure 7(b) plots pressure $P$ and number density $n$ at $t$ = 120 Myrs. Since the warm gas infalling along the closed magnetic field lines compresses the gas around the bottom of the flux rope ($x$ = 0 pc), the number density
exceeds the threshold for the cooling instability, and deviates from the warm gas branch. The dense, cool
gas is transformed to the cold neutral medium (CNM) where temperature is less than 400 K. Since we cut off cooling when $T <$ 400 K to numerically resolve the filament, the CNM locates at the isothermal line
at $T$ = 400 K.   

Figure 8 shows time evolution of density, pressure and temperature at $x$ = 0 pc and $z$ = 105 pc. The initially warm gas ($T \sim$ 10$^4$ K) is compressed during 30-50 Myrs. Consequently, density and pressure increase and temperature slightly decreases. At $t$ = 50 Myrs, the gas locates in the thermally unstable region in the $P$-$n$ diagram. Figure 8 shows that pressure and temperature decrease due to the cooling instability and the density continues to increase. Finally, cool ($T$ = 400 K), dense filament is formed.

Figure 9 shows vertical distribution of temperature at $x$ = 0 pc at 120 Myrs. The dense filament locates at $z$ = 100-160 pc. Meanwhile, high temperature ($T$ = 35000 K-70000 K) region appears at $z$ = 330-450 pc. The hot region is formed by shock waves formed by the upward flow.
When the temperature of the shocked gas exceeds 20000 K, the region stays in high temperature state.

Figure 10 shows column density of the dense gas ($T <$ 500 K), the unstable neutral medium (UNM) in the temperature range 500 K $< T <$ 5000 K and the warm gas in the temperature range 5000 K $< T <$ 7000 K, at 120 Myrs for line of sight direction parallel to the $x$-axis. 
Column density of the dense filament is $\sim$ 1.0 $\times$ 10$^{-3}$ g cm$^{-2}$ (column number density = 6 $\times$ 10$^{20}$ cm$^{-2}$) around the bottom of the dense filament. The column density of the unstable neutral medium is at largest 1.0 $\times$ 10$^{-4}$ g cm$^{-2}$ (column number density = 6 $\times$ 10$^{19}$ cm$^{-2}$). The column density of the warm gas is 3 $\times$ 10$^{-4}$ g cm$^{-2}$ above the top of the dense filament. 
The dense gas can be observed by CO emission and the UNM and the warm gas can be observed by 21cm line emission by neutral hydrogen. \citet{2010PASJ...62.1307T} compared the distribution of CO emission and HI emission and found that they coincide in loop 1. In our simulation, the distribution of the UNM coincides with the dense filament but the warm gas above the dense filament has comparable column density as the dense filament. 

   The column density estimated from the CO emission is 3 $\times$ 10$^{21}$ cm$^{-2}$ around the top of the loop 1 and the HI column number density of loop 1 obtained from the 21cm line emission is 6 $\times$ 10$^{20}$ cm$^{-2}$ \citep{2010PASJ...62.1307T}. These column densities are 5 times larger than that of our simulation, and will be discussed in the next subsection.

\begin{figure}
\figurenum{4}
\gridline{\fig{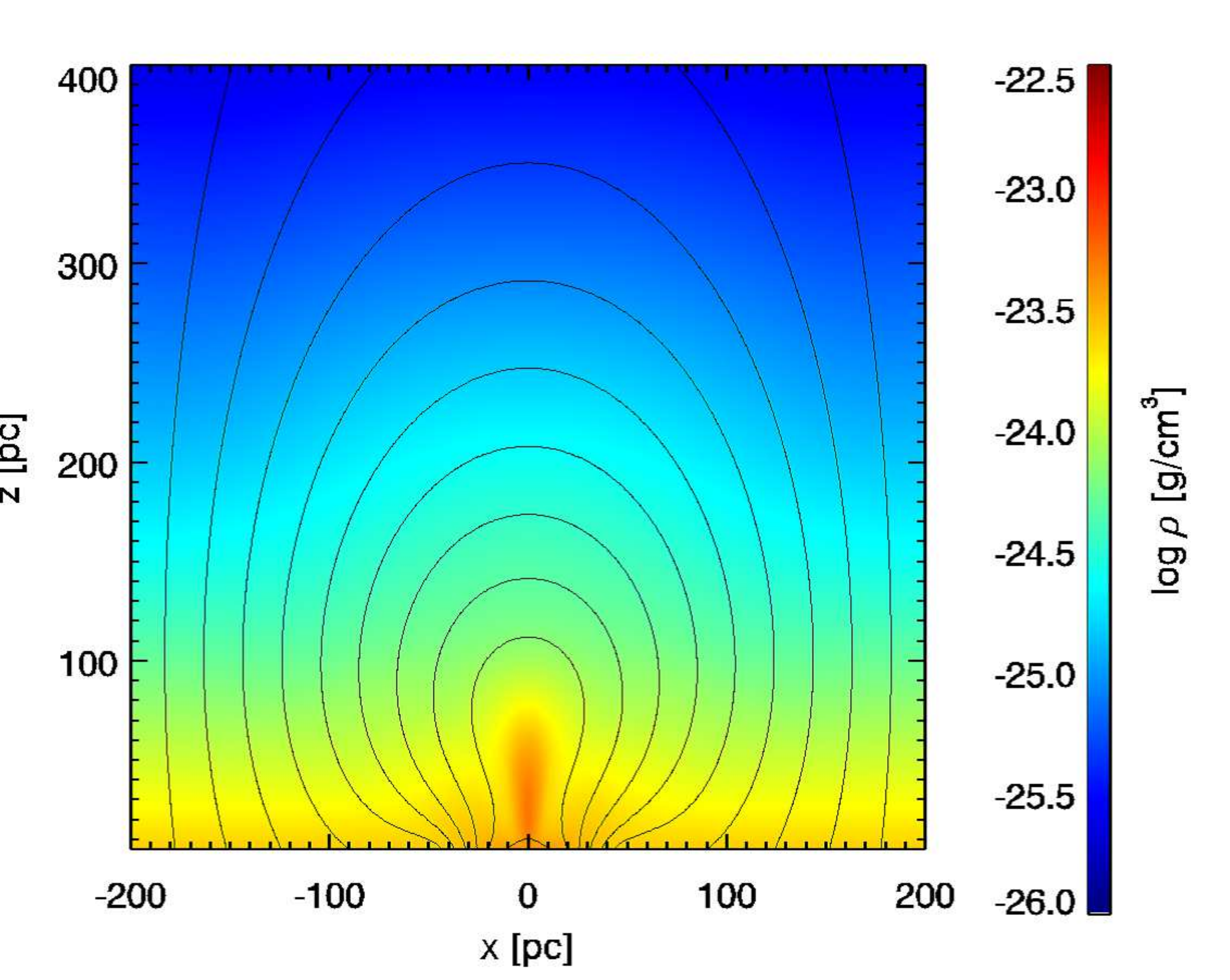}{0.45\textwidth}{Time = 30 Myr}
          \fig{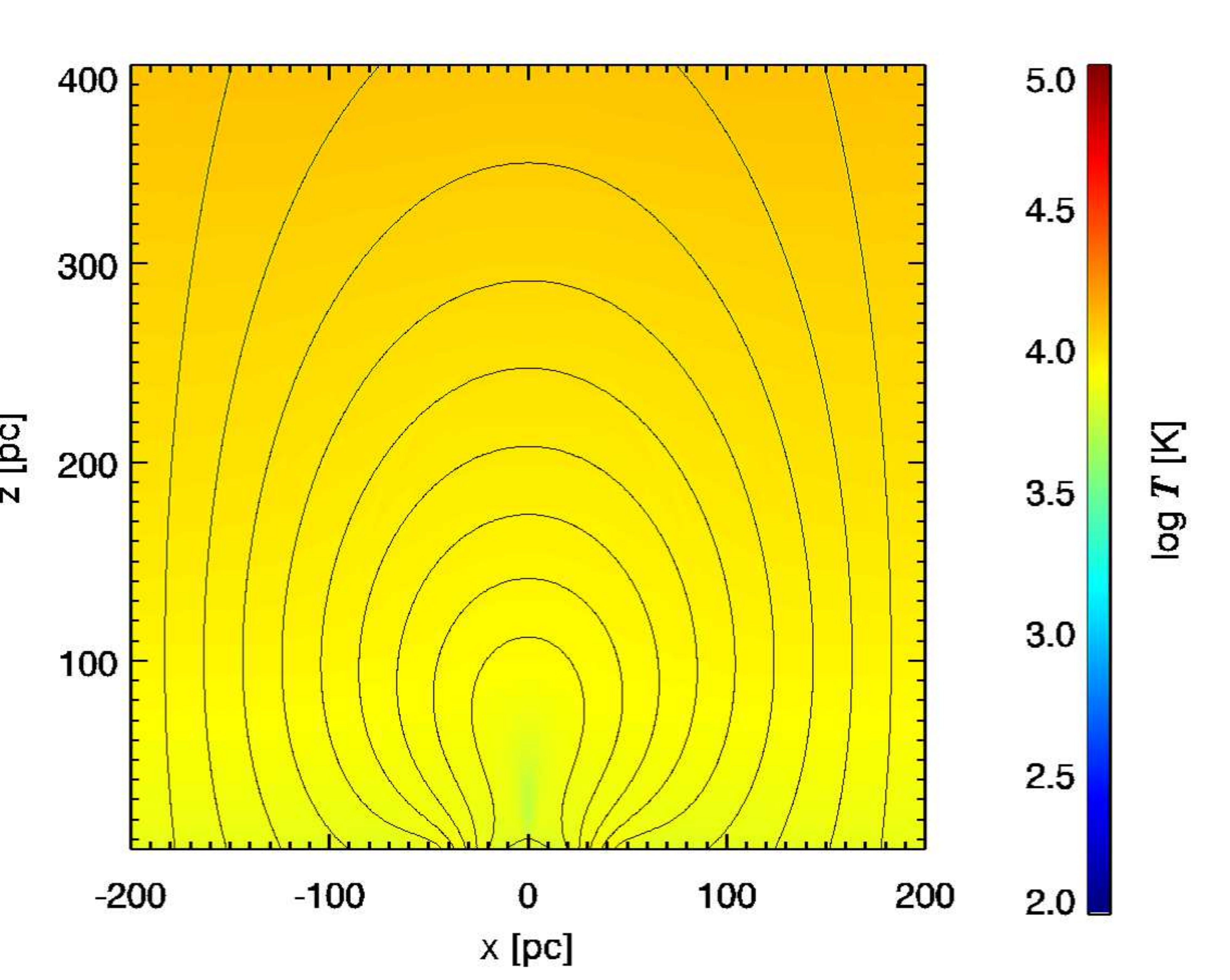}{0.45\textwidth}{Time = 30 Myr}
          }
\gridline{\fig{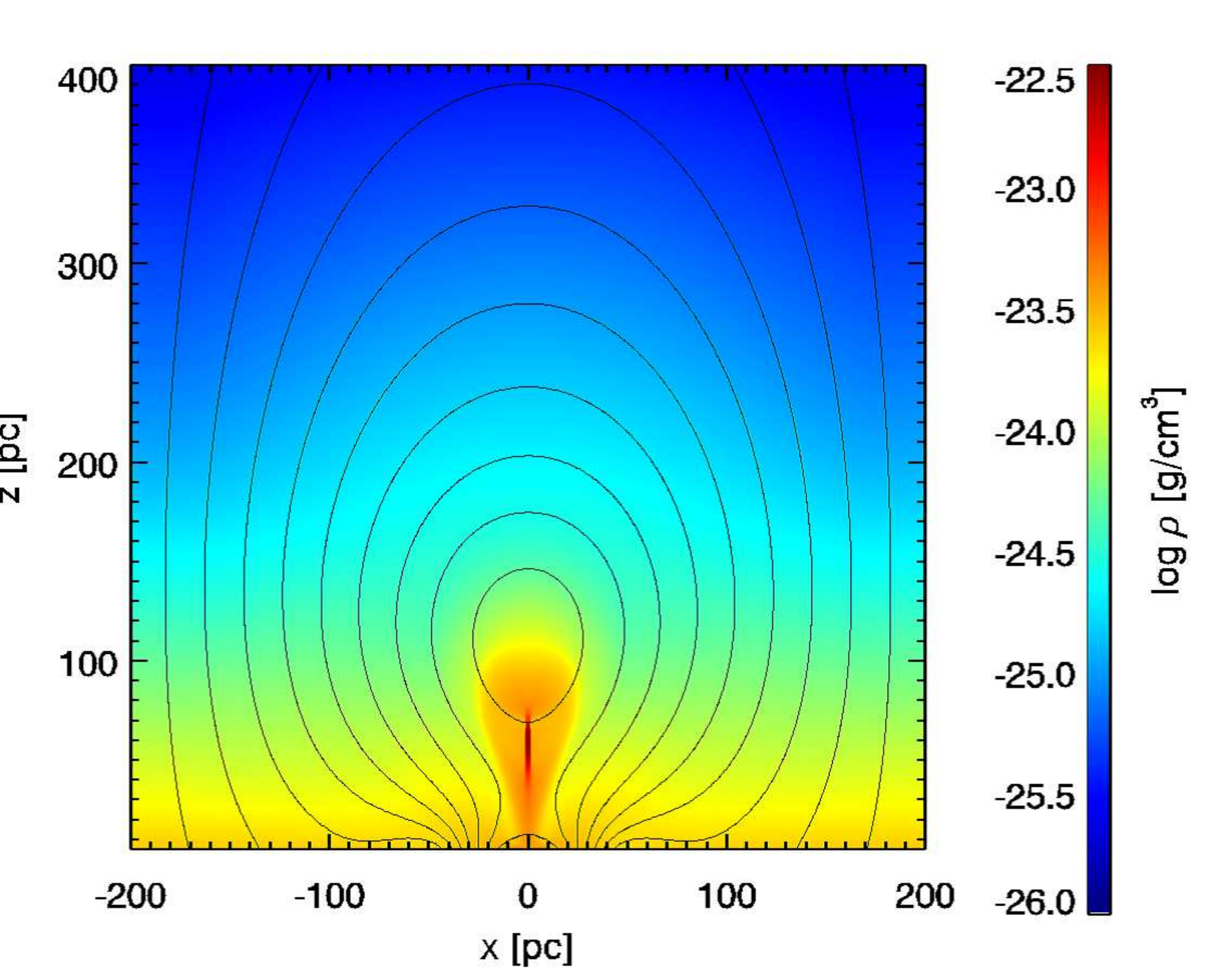}{0.45\textwidth}{Time = 40 Myr}
          \fig{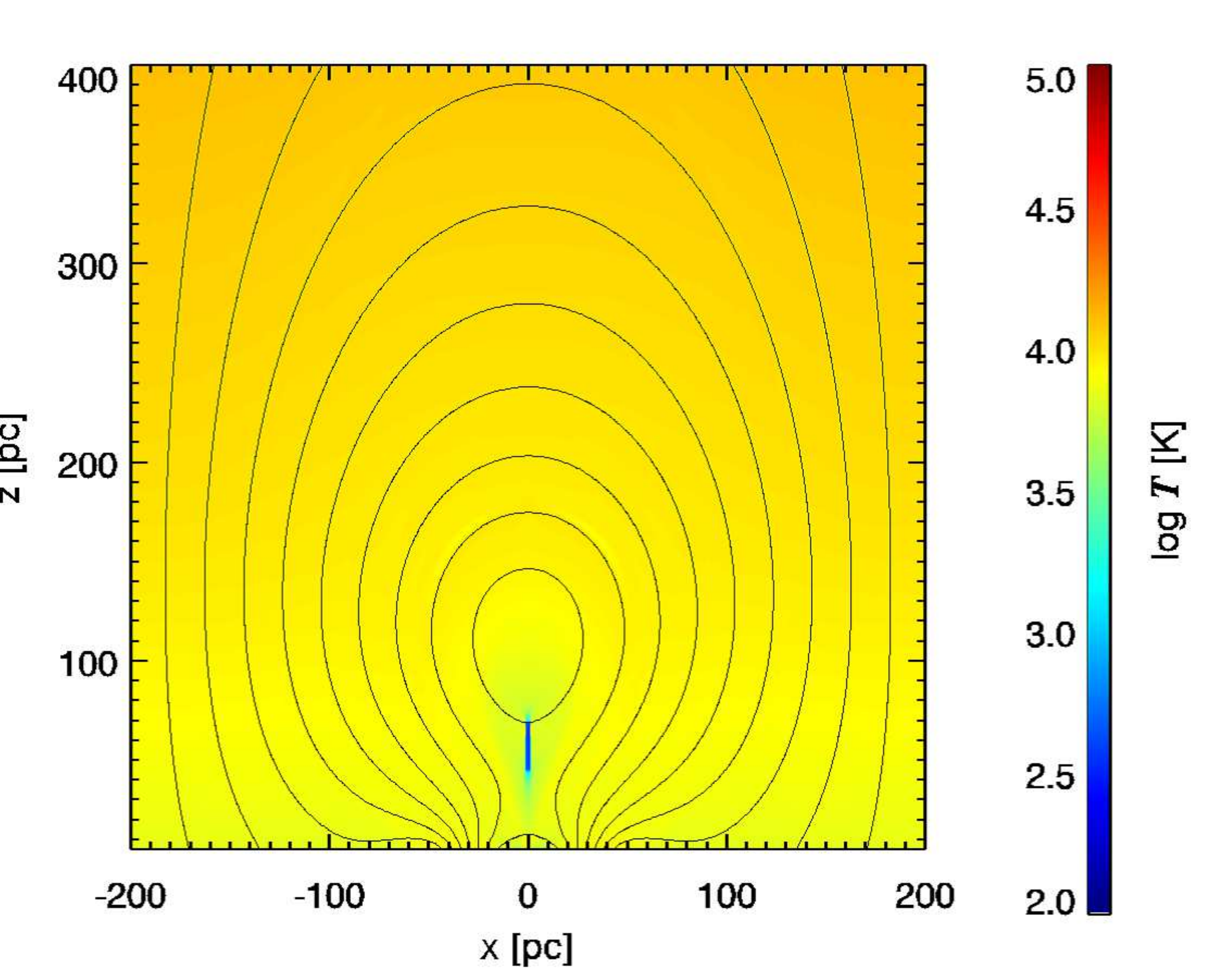}{0.45\textwidth}{Time = 40 Myr}
          }
\gridline{\fig{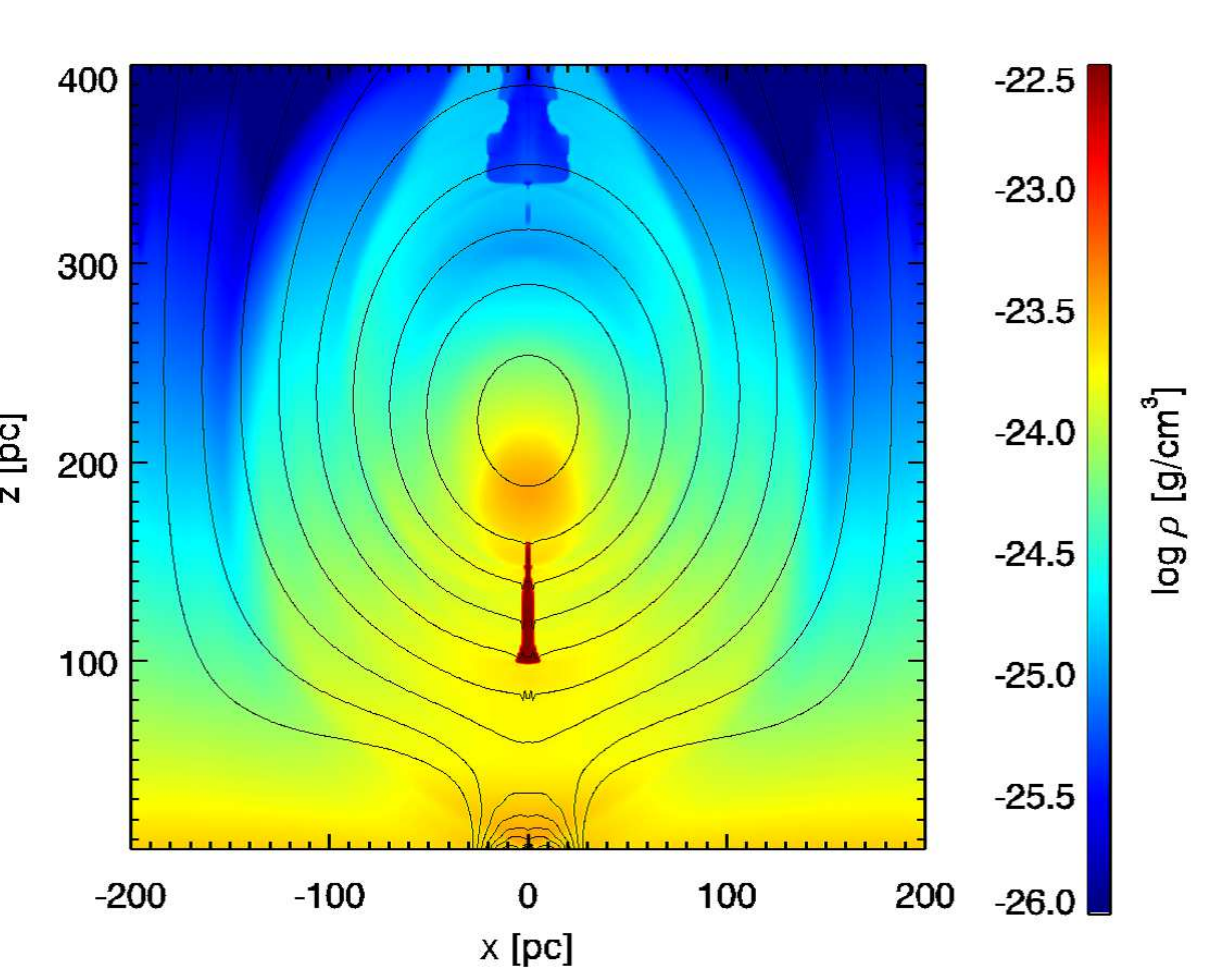}{0.45\textwidth}{Time = 120 Myr}
          \fig{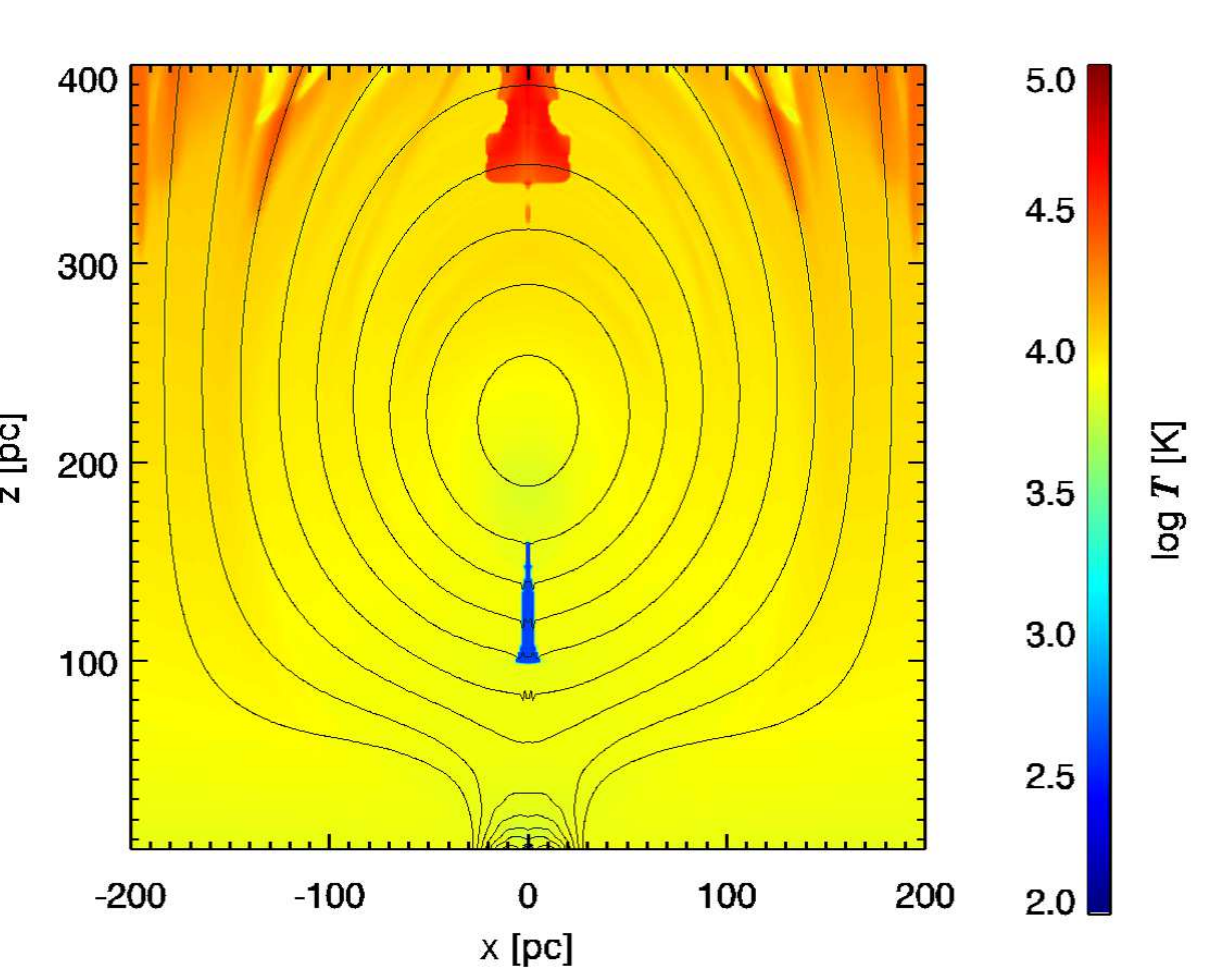}{0.45\textwidth}{Time = 120 Myr}
          }
\caption{Density (left panels), temperature (right panels) distribution and magnetic field lines (solid lines) at 30 Myrs (top), 40 Myrs (middle) and 120 Myrs (bottom).\label{fig:pyramid}}
\end{figure}

\begin{figure}
\figurenum{5}
\epsscale{0.85}
\plotone{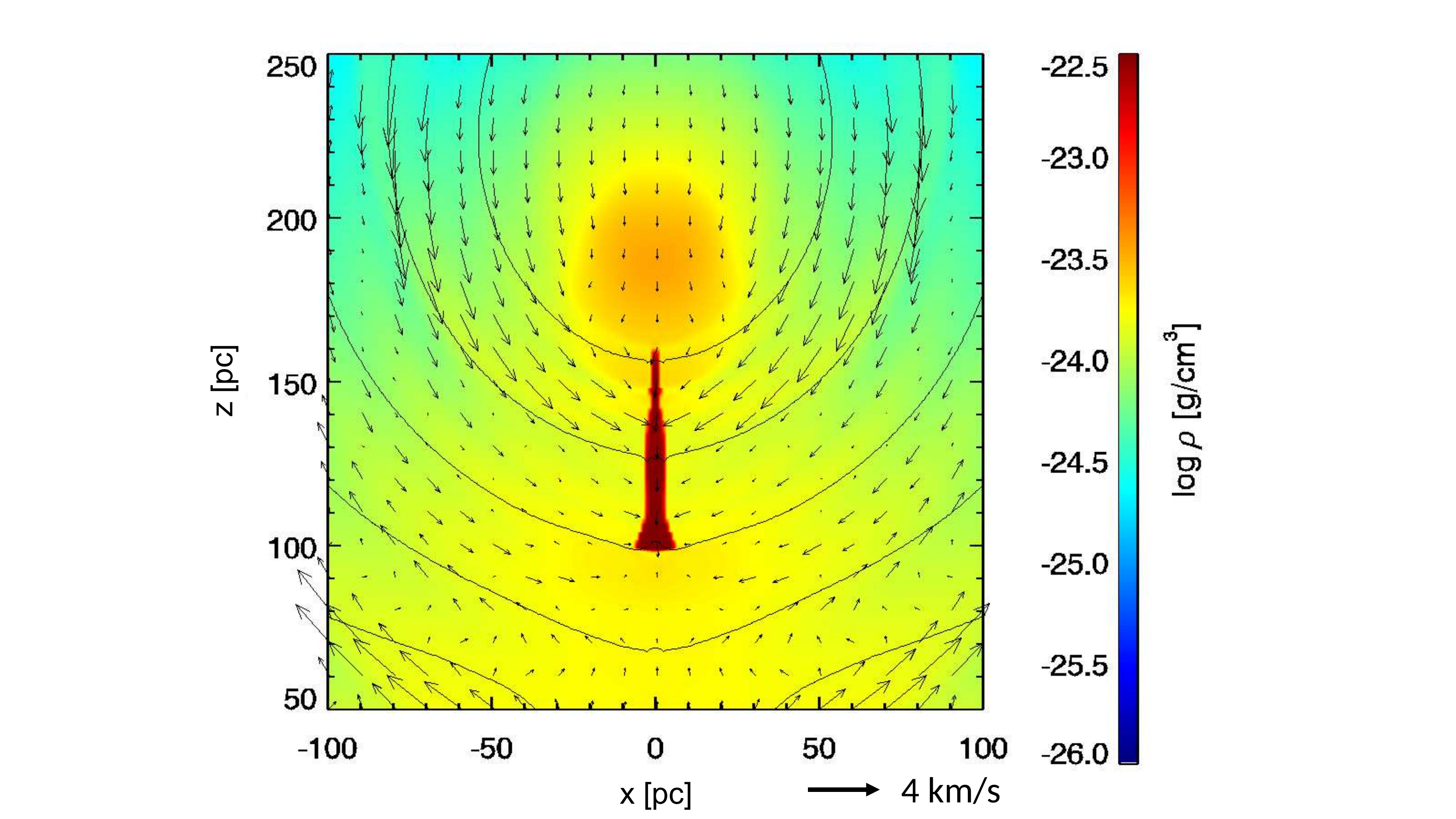}
\caption{Density distribution (color), magnetic field lines (solid line) and velocities (arrows).}
\end{figure}

\begin{figure}
\figurenum{6}
\epsscale{0.6}
\plotone{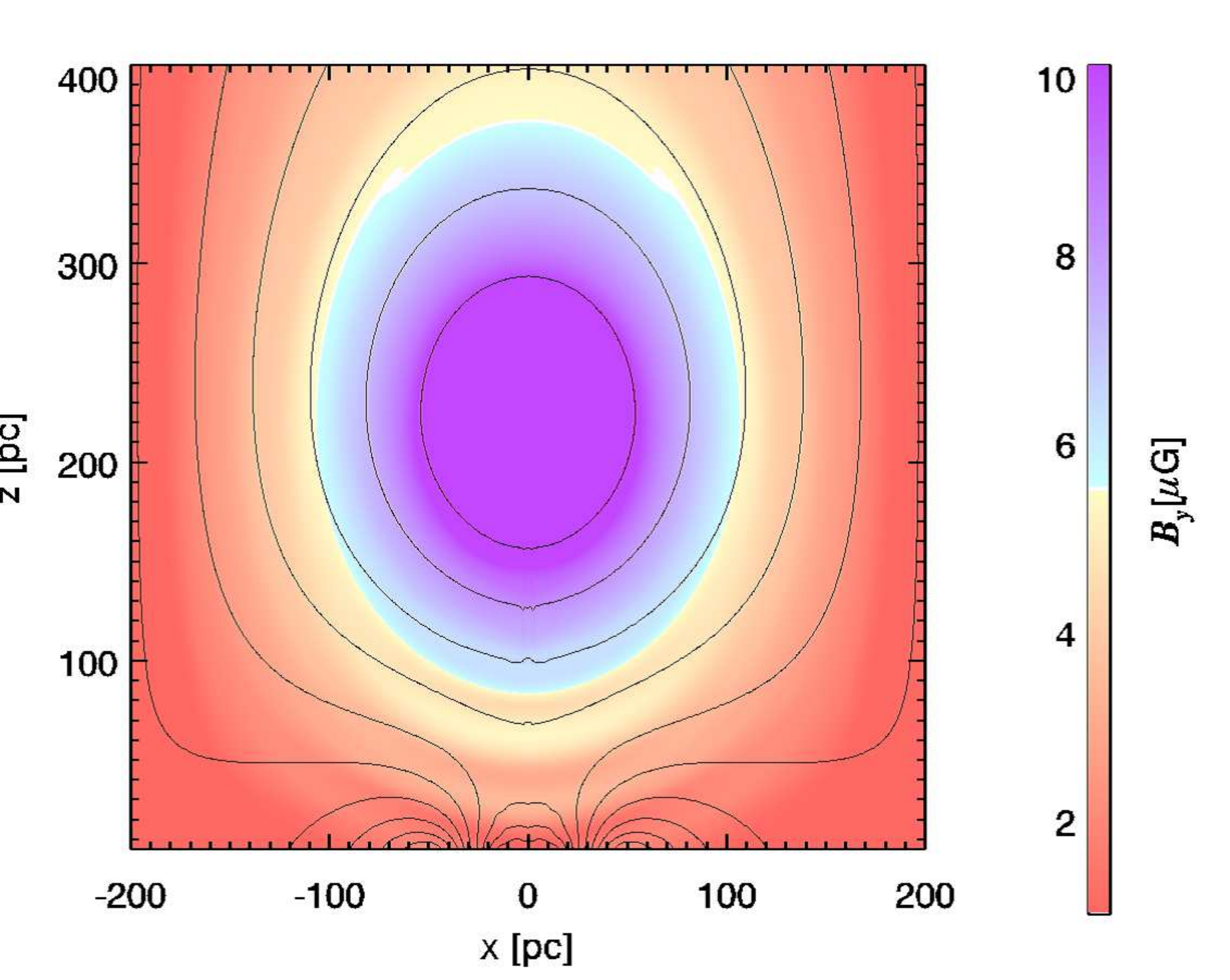}
\caption{Distribution of $B_y$ and magnetic field lines (solid line).}
\end{figure}

\begin{figure}
\figurenum{7}
\gridline{\fig{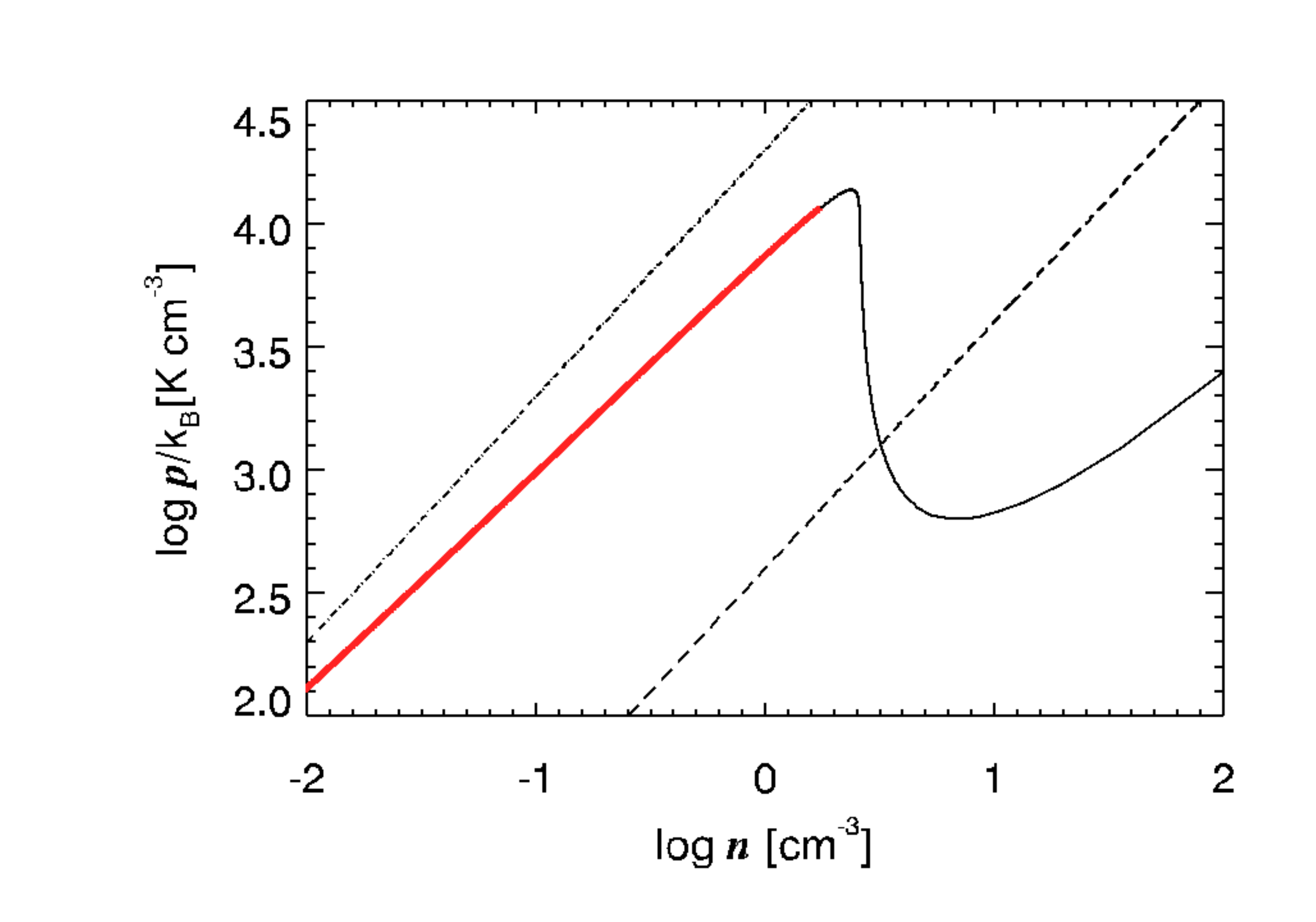}{0.5\textwidth}{Time = 0 Myr}
          \fig{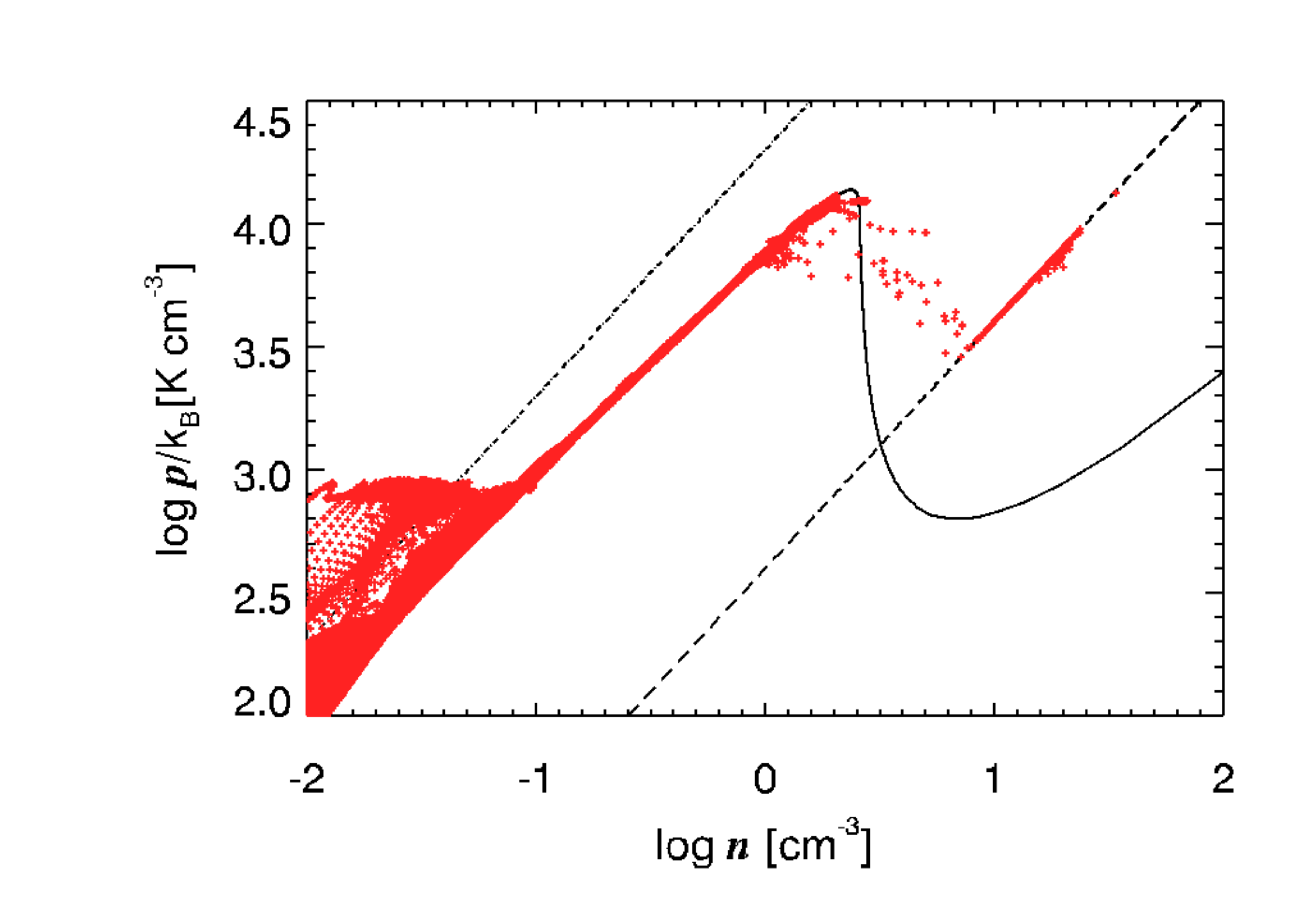}{0.5\textwidth}{Time = 120 Myr}
          }          
\caption{Pressure and number density at 0 Myr (left panels) and 120 Myrs (right panels).Solid curve shows thermal equilibrium curve. The dashed line shows the isothermal line at $T$ = 400 K and the dash-dotted line shows that at $T$ = 20000 K.}
\end{figure}

\begin{figure}
\figurenum{8}
\epsscale{0.8}
\plotone{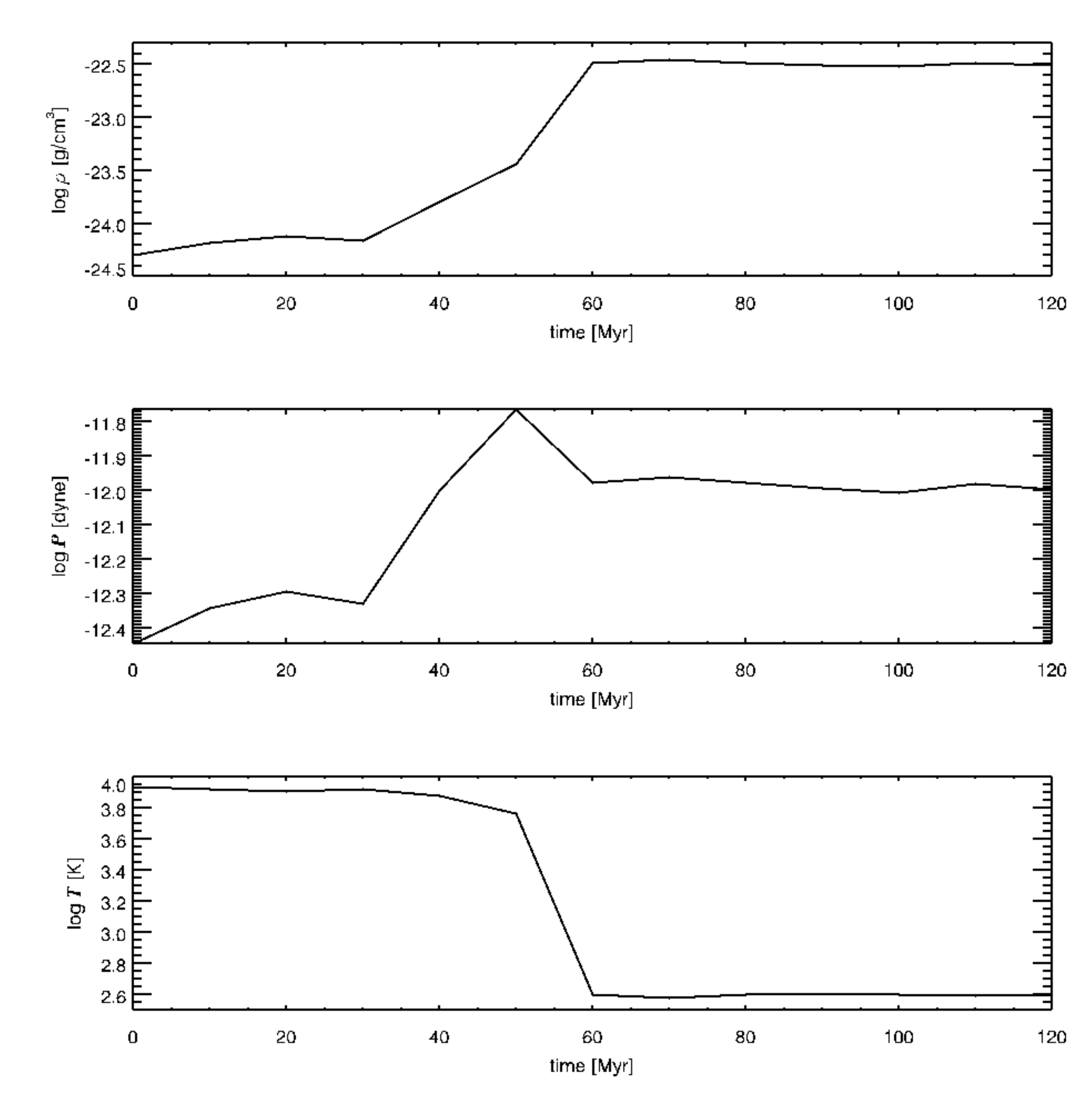}
\caption{Time evolution of density (top), pressure (middle) and temperature (bottom) at $x$ = 0 pc and $z$ = 105 pc. }
\end{figure}

\begin{figure}
\figurenum{9}
\epsscale{0.9}
\plotone{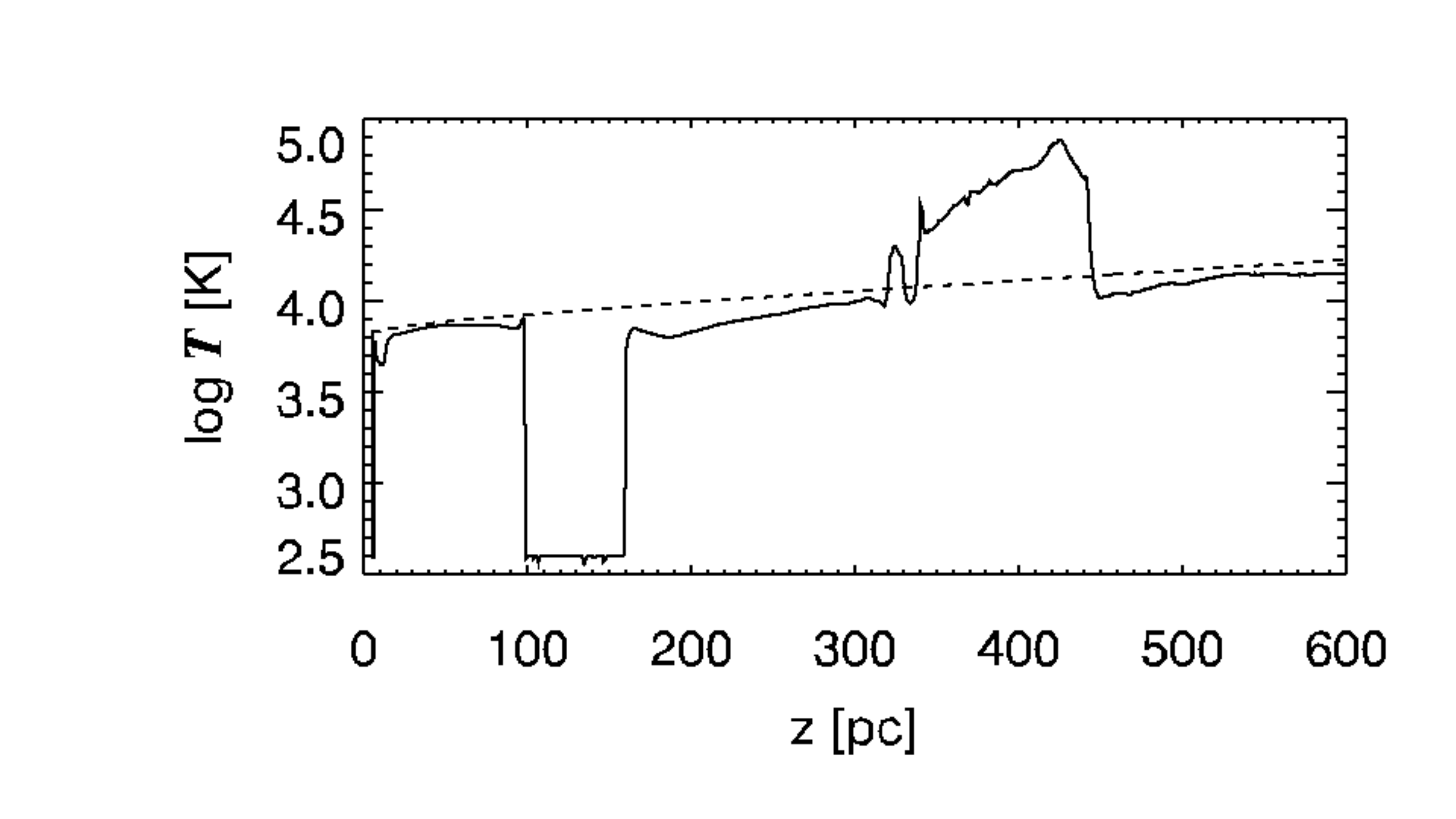}
\caption{Vertical distribution of temperature at $x$ = 0 pc at 120 Myrs (solid line) and initial temperature profile (dashed line).}
\end{figure}

\begin{figure}
\figurenum{10}
\epsscale{0.8}
\plotone{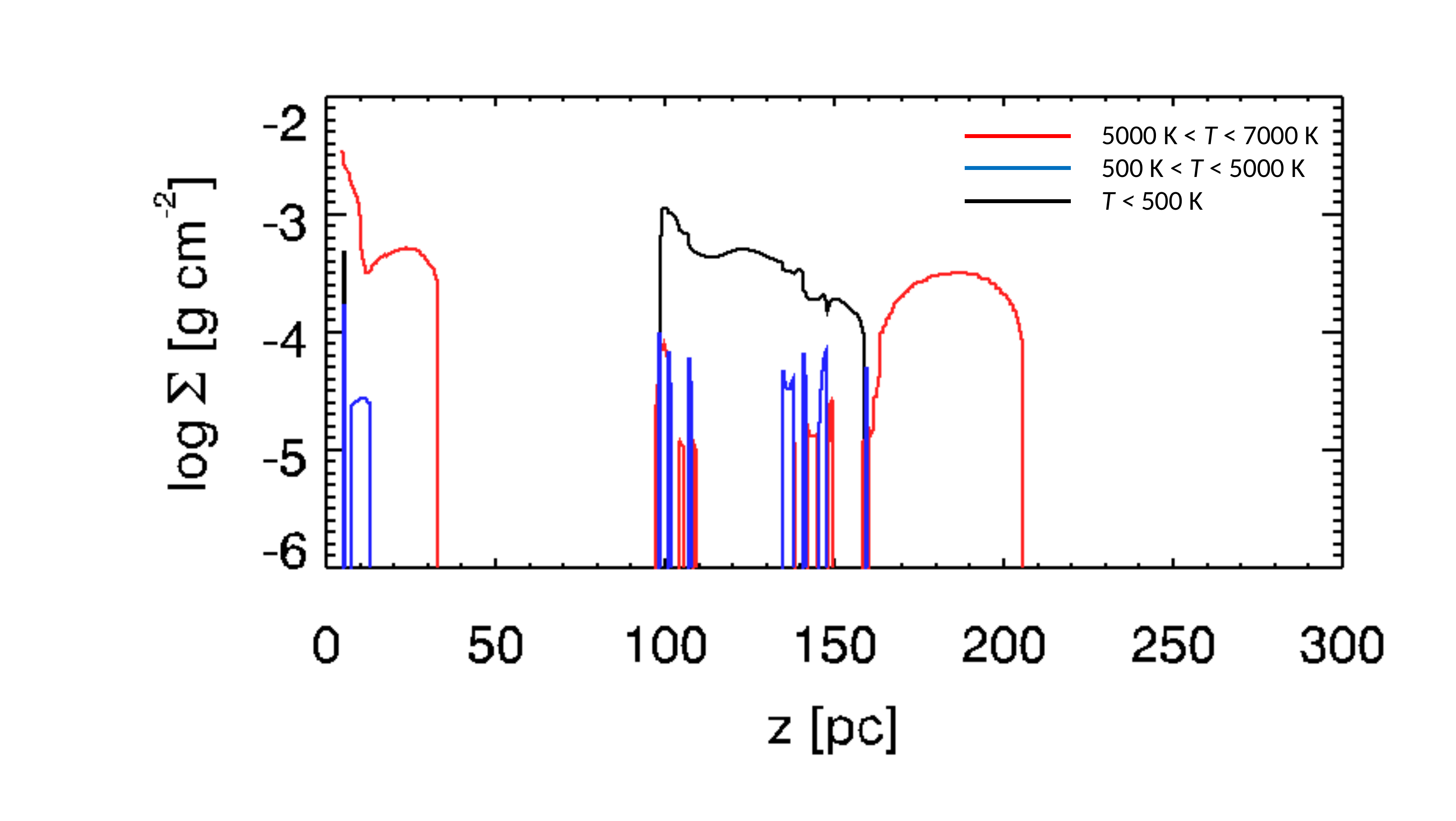}
\caption{Vertical distribution of column density of the dense gas ($T <$ 500 K) with black line and the unstable neutral gas (500 K $< T <$ 5000 K) and warm gas (5000 K $< T <$ 7000 K) with blue and red curves, respectively. The line of sight direction is parallel to the $x$-axis. }
\end{figure}

\subsection{Mass of The Filament}
In this section, we would like to estimate the mass levitated in the cold, dense filament and compare the numerical results with observations.
NANTEN observations indicate that the projected lengths of two loops (loop 1 and loop 2) are $\sim$500 pc and $\sim$300 pc, respectively. They are lifted up to $\sim$220 pc and $\sim$300 pc and total mass of each loop was estimated to be 8 $\times $ 10${^4}$ M$_{\odot}$ \citep{2006Sci...314..106F}. If we take average value of projected lengths of two loops and assume that the shape of loops are a half circle, the average length of the loop is 400 pc $\times$ $\pi$/2, $\sim$ 600 pc. In our simulation results at 120 Myrs, dense filament ($T <$ 500 K) is lifted up to 100$\sim$200 pc. The mass of the filament can be estimated to be 112 M$_{\odot}$ times the length of the loop in $y$-direction in unit of pc. In other words, the total mass of the dense filament is 7 $\times$ 10$^4$ M$_{\odot}$ if we assume the length of the filament in $y$-direction is 600 pc. 

The total mass of the dense filament in the simulation is consistent with the lower limit of the mass estimated from observation. However, \citet{2010PASJ...62.1307T} estimated by detailed analysis of the NANTEN $^{12}$CO and $^{13}$CO($J$ = 1-0) observations that the total mass of these two loops are 1.4 $\times$ 10$^{6}$ M$_{\odot}$ and 1.9 $\times$ 10$^{6}$ M$_{\odot}$, respectively. These values are an order of magnitude larger than those of our simulation.
The smaller total mass of the molecular loop in our simulation can be due to the lower density of the warm medium (WM), $\rho_{WM}$ = 2.8 $\times$ 10$^{-24}$  g cm$^{-3}$ we assumed at the bottom of our simulation box at $z$ = 5 pc. 
One possibility is that the heating rate $\Gamma$ is larger
   in the Galactic central region.    
   When $\Gamma_{GC}$ = 10$^{-25}$ erg s$^{-1}$, since the thermal equilibrium
   condition gives $\rho_{WM}$($z$ = 5 pc, GC) $\sim$ 1.4 $\times$ 10$^{-23}$ g cm$^{-3}$,
   the total mass of the filament can be 5 times that for
   $\rho_{WM}$($z$ = 5 pc) = 2.8 $\times$ 10$^{-24}$ g cm$^{-3}$.
   Another possibility is the co-existence of the molecular
   gas with warm gas at the initial state.
   According to \citet{2016PASJ...68...63S}, the fraction of molecular gas
   $f_{mol}$ = $\rho_{H2}V_{H2}$/($\rho_{WM}V_{WM}$ + $\rho_{H2}V_{H2}$) $\sim$ 0.8 at z $\sim$
   20 pc in the Galactic central region where
   $\rho_{H2}$ is the density of molecular hydrogen, and $V_{WM}$ and $V_{H2}$
   are volume fraction of the warm gas and molecular hydrogen,
   respectively. It means that when $\rho_{WM}$ = 2.8 $\times$ 10$^{-24}$ g cm$^{-3}$
   and $\rho_{H2}$ = 2.8 $\times$ 10$^{-22}$ g cm$^{-3}$, the volume fraction of
   the molecular hydrogen $V_{H2}$ $\sim$ 0.04.
   Since the mean density of the two phase medium is
   $\rho$ = $\rho_{WM}V_{WM}$ + $\rho_{H2}V_{H2}$ = 1.4 $\times$ 10$^{-23}$ g cm$^{-3}$,
   the total mass of the filament can be 5 times that for the
   warm medium with $\rho_{WM}$ = 2.8 $\times$ 10$^{-24}$ g cm$^{-3}$.
   The third possibility is the larger size of the magnetic arch.
   We assumed the half width of the magnetic arch is 200 pc.
   When the magnetic arch is larger, it forms larger flux rope
   and more mass inside the rope accumulates, so that the total
   mass of the dense filament can be larger.

Here we would like to point out that the mass of the dense filament increases with time. Figure 11 shows the time evolution of the total mass of the dense filament. As we have shown in Figure 5, warm gas in the magnetic flux rope continuously fall down towards the filament along magnetic field lines. The maximum mass of the filament can be estimated by assuming that all the gas inside a flux rope fall into the filament. The mass inside the flux rope corresponding to the blue-purple region in Figure 6 is 3.5 $\times$ 10$^{5}$ M$_{\odot}$ assuming that the depth in $y$-direction is 600 pc.

\begin{figure}
\figurenum{11}
\epsscale{0.8}
\plotone{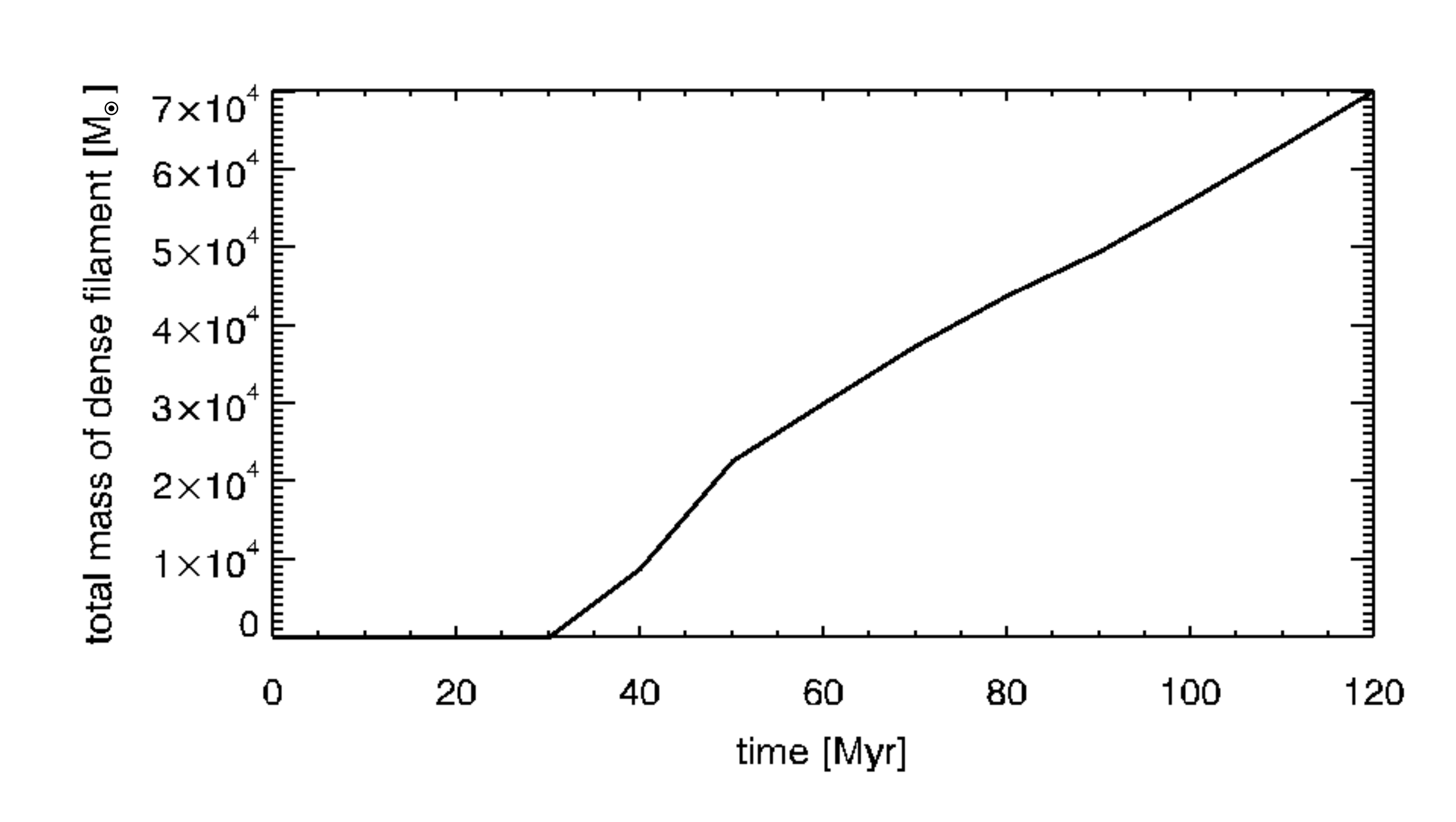}
\caption{Time evolution of total mass of a dense filament.}
\end{figure}



\section{Summary and Discussion} \label{sec:discussion}
The idea of "Galactic Prominence" was addressed in \citet{2006Sci...314...70M} and \citet{2010PASJ...62.1307T} to explain the formation mechanism of the molecular loops found in the Galactic central region. The essential component which sustains the prominence is strong magnetic field, which enable the dense filament to be levitated against gravity. In \citet{2006Sci...314..106F}, the observed velocity gradient along the loop, and large velocity dispersion around the footpoint of the loop were explained by the sliding motion of the molecular gas along the magnetic loop formed by the Parker instability. However, since the dense molecular gas in the equatorial region are hard to be lifted by magnetic buoyancy, the molecular gas should be formed in-situ in the rising magnetic loops. A possible mechanism of the in-situ formation of molecular gas is compression of the warm neutral medium by magnetic loops buoyantly rising by Parker instability. Magnetohydrodynamic simulations by \citet{2009PASJ...61..957T}, however, showed that the maximum density of the dense shell formed around the top of the rising magnetic loops is too small to trigger the cooling instability. We may need preexisting dense gas layer to trigger the cooling instability. In this paper, we applied the in-situ formation model of solar prominences by \citet{2015ApJ...806..115K} to the Galactic gas disk and demonstrated that dense, cold filaments can be formed by the cooling instability. Since we assumed translational symmetry to simulate the prominence formation by 2.5D MHD simulation, we cannot reproduce the velocity gradient along the loop, which needs 3D simulation.

Recently, \textit{ESA's Herschel space observatory} carried out detailed observations of dense filaments in our Milky Way \citep{2013A&A...550A..38P,2014ApJ...791...27S,2015MNRAS.450.4043W}. Filaments whose length ranges form a few pc to hundred pc are found along the galactic plane. Three-dimensional magnetohydrodynamic simulation by \citet{2016MNRAS.459.1803W} demonstrated that the dense filaments can be formed by thermal instability in magnetized interstellar medium without assuming colliding flows or other external effects. Filaments with higher aspect ratio can be formed by stronger magnetic field.


In our simulation, the total mass of the dense filament is 7 $\times$ 10$^4$ M$_\odot$ at t = 120 Myrs. It is smaller than 1.6 $\times$ 10$^6$ M$_\odot$ observed for loops 1 by NANTEN. The total mass of the filament can be larger if we take into account the enhanced heating in the Galactic center and the co-existence of the molecular gas with warm medium. The mass of the filament can be larger when the size of the magnetic arcade is larger.

In this paper, we neglected cooling when $T <$ 400K. If we include cooling in this temperature range, the dense filaments can be thinner. We will need more grid points to resolve such filaments. Since the thickness of the filament becomes less than the Field's length, we may need to 
include thermal conduction. 
According to \citet{2006ApJ...652.1331I}, the thermal conductivity drives turbulence in two phase interstellar medium if we resolve the Field's length of the warm gas ($\sim$ 0.1 pc). Such turbulence may thicken the dense filament, and enable us to compare the numerical results more quantitatively with observations.
We would like to report the results of such simulations in subsequent papers.

\acknowledgments
We appreciate discussion and helpful comments by  M. Machida, T. Hanawa, T. Kaneko, Y. Kudoh and Y. Matsumoto. The simulation codes we adopted in our simulation is developed by Y. Asahina. C.-H. Peng appreciate the scholarship from Interchange Association, Japan. Numerical computations were carried out on Cray XC30 at Center for Computational Astrophysics, National Astronomical Observatory of Japan.



\vspace{5mm}
\listofchanges

\end{document}